\documentclass[12pt]{article}
\usepackage{amssymb,amsmath,epsfig}
\allowdisplaybreaks

\begin{document}
\title{\bf Electromagnetic Effects on Cracking of Anisotropic Polytropes}
\author{M. Sharif \thanks{msharif.math@pu.edu.pk} and Sobia Sadiq
\thanks{sobiasadiq.01@gmail.com}\\
Department of Mathematics, University of the Punjab,\\
Quaid-e-Azam Campus, Lahore-54590, Pakistan.}

\date{}
\maketitle
\begin{abstract}
In this paper, we study the electromagnetic effects on stability of
spherically symmetric anisotropic fluid distribution satisfying two
polytropic equations of state and construct the corresponding
generalized Tolman Oppenheimer Volkoff equations. We apply
perturbations on matter variables via polytropic constant as well as
polytropic index and formulate the force distribution function. It
is found that the compact object is stable for feasible choice of
perturbed polytropic index in the presence of charge.
\end{abstract}

\section{Introduction}

A stellar object is worthless if it is not stable towards
perturbations in its physical variables (e.g. energy density and
pressure anisotropy). The stability analysis of such objects is an
interesting issue in general relativity and astrophysics. Bondi
\cite{1} was the pioneer to develop hydrostatic equilibrium equation
to examine the stability of self-gravitating spheres. Herrera
\cite{2} introduced the concept of cracking as well as overturning
to describe the behavior of isotropic and anisotropic matter
distribution just after its deviation from equilibrium state. The
cracking is observed when the radial force is positive in the inner
regions but negative at outer ones while overturning is produced for
the reverse situation. In general, cracking and overturning occur
when total radial forces change their signs within the matter
distribution.

In stellar objects, pressure anisotropy is an important matter
ingredient which affects their evolution and can be produced via
different physical processes such as phase transition and mixture of
two fluids etc \cite{2aa}. Bowers and Liang \cite{3} studied static
spherically symmetric anisotropic matter configuration and found an
increase in surface redshift as well as equilibrium mass of the
system. Gokhroo and Mehra \cite{5} obtained solutions of anisotropic
Einstein field equations by considering variable energy density and
observed larger redshifts of spherical objects. Mak and Harko
\cite{4} discussed static anisotropic spherical stars and concluded
that energy density, radial and tangential pressures are positive in
the interior regions. Cosenza et al. \cite{6} found solutions of the
field equations with spherically symmetric anisotropic matter
distribution.

The polytropic equation of state (EoS) has captivated the attention
of many researchers to discuss the internal structure of compact
objects. It is a power-law relationship between energy density and
pressure defined as
\begin{equation}\nonumber
P=k\rho^\gamma=k\rho^{1+\frac{1}{n}},
\end{equation}
where $P,~\rho,~\gamma,~k$ and $n$ denote pressure, energy density,
polytropic exponent, polytropic constant and polytropic index,
respectively. Tooper \cite{7} gave the idea of relativistic study of
polytropes and formulated two non-linear differential equations
describing the stellar structure. He found physical variables (mass,
pressure and density) of polytropes using numerical technique.
Herrera and Barreto \cite{8} discussed general formalism for
relativistic isotropic as well as anisotropic polytropes of
spherically symmetric matter distribution and constructed the
Lane-Emden equation which represents the inner structure of compact
objects. Herrera et al. \cite{9} examined the stability of
spherically symmetric anisotropic conformally flat polytropes using
Tolman mass and found that the considered polytropic models are
stable.

Perturbation technique plays a crucial role in the stability of
astrophysical objects. A star can collapse, expand, crack or
overturn depending upon the nature of perturbation. Di Prisco et al.
\cite{b} discussed the cracking of spherical compact object and
concluded that departure from equilibrium state leads to the
cracking only for local (non-constant) anisotropic perturbation.
Abreu et al. \cite{a} examined the impact of energy density as well
as local anisotropic perturbation on the stability of local and
non-local anisotropic matter distributions. They found that for
$P_{r}=0,~P_{\bot}\neq0$, the matter configuration is always stable
whereas it can experience a cracking (or overturning) when
$P_{\bot}=0~ \text{and}~ P_{r}\neq0$. Gonzalez et al. \cite{d}
investigated the effect of local density perturbations for isotropic
as well as anisotropic spheres satisfying barotropic EoS. They
obtained that isotropic configuration also presents cracking which
is in contrast to the result obtained for non-local density
perturbation. Recently, Herrera et al. \cite{c} observed the
cracking and overturning for anisotropic polytropes by perturbing
energy density as well as local anisotropy of the system.

The study of charge in self-gravitating spherically symmetric matter
distributions started with the contributions of Rosseland and
Eddington \cite{1a}. Since then numerous efforts have been made to
explore the effects of charge on the structure and evolution of
self-gravitating systems. Bekenstein \cite{1s} discussed the
gravitational collapse of charged spherically symmetric perfect
fluid by introducing the idea of hydrostatic equilibrium. Bonnor
\cite{2s} investigated the role of charge in the collapse of
spherical dust cloud and found that it halts the process of
collapse. Ray et al. \cite{12} examined the effects of charge on
compact stars and obtained $10^{10}$ coulumbs charge present in high
density compact objects. Takisa and Maharaj \cite{13} formulated
exact solutions of Einstein-Maxwell field equations with polytropic
EoS which can be used to model charged anisotropic compact objects.
We have studied the modeling of charged conformally flat polytropic
sphere and checked their viability through energy conditions
\cite{3s}. Azam et al. \cite{4s} examined the occurrence of cracking
in charged static spherically symmetric compact objects with
quadratic EoS and concluded that stability regions increase by the
increase of charge.

In this paper, we study the cracking of spherically symmetric
anisotropic polytropes in the presence of charge. The format of this
paper is as follows. In the next section, we study matter
distribution for charged sphere and construct the generalized Tolman
Oppenheimer Volkoff (TOV) equation as well as mass equation for two
cases of polytropic EoS. Section \textbf{3} is devoted to observe
the cracking by perturbing energy density and local anisotropy via
polytropic parameters. We formulate force distribution function for
each case and plot the results numerically to discuss stability of
corresponding models. Finally, we conclude our results in the last
section.

\section{Matter Distribution and Generalized TOV Equation}

We consider static spherically symmetric spacetime as follows
\begin{equation}\label{1}
ds^2=-e^{\nu(r)} dt^{2}+e^{\lambda(r)}
dr^{2}+r^{2}\left(d\theta^{2}+\sin^{2}\theta d\phi^2\right).
\end{equation}
The matter distribution is considered to be anisotropic in pressure
bounded by hypersurface $\Sigma$ so that
$r=r_{\Sigma}=\textmd{constant}$. The energy-momentum tensor for
such matter distribution is given by
\begin{equation}\label{2}
T_{\mu\nu}=(\rho+P_{\bot})V_\mu V_\nu+P_\bot g_{\mu\nu} + (P_r -
P_\bot)S_\mu S_\nu,
\end{equation}
where $P_{r},~P_{\perp},~V_{\mu},~S_\mu$ are the radial pressure,
tangential pressure, four-velocity and four-vector, respectively. We
consider fluid to be comoving as
\begin{equation*}
S^{\mu}=e^{-\lambda/2}\delta^{\mu}_{1},\quad
V^{\mu}=e^{-\nu/2}\delta^{\mu}_{0},
\end{equation*}
satisfying
\begin{equation*}
V^{\mu}V_{\mu}=-1,\quad S^{\mu}S_{\mu}=1, \quad S^{\mu}V_{\mu}=0.
\end{equation*}
The energy-momentum tensor for electromagnetic field is defined as
\begin{equation}\label{3}
E_{\mu\nu}=\frac{1}{4\pi}\left(F_{\mu}~^{\alpha}F_{\nu\alpha}-
\frac{1}{4}F^{\alpha\beta}F_{\alpha\beta}g_{\mu\nu}\right),
\end{equation}
where $F_{\mu\nu}=\phi_{\nu,\mu}-\phi_{\mu,\nu}$ and $\phi_\mu$ are
the Maxwell field tensor and four potential, respectively. The
Maxwell field tensor satisfies the following field equations
\begin{equation*}
F^{\mu\nu}_{~~;\nu}={\mu}_{0}J^{\mu},\quad F_{[\mu\nu;\gamma]}=0,
\end{equation*}
here $\mu_0$ is the magnetic permeability and $J^{\mu}$ is the four
current. In comoving coordinates, we have
\begin{equation*}
\phi_{\mu}={\phi} {\delta^{0}_{\mu}},\quad J_{\mu}=\zeta V_{\mu},
\end{equation*}
where $\zeta=\zeta(r)$ and $\phi=\phi(r)$ represent scalar potential
and charge density, respectively.

The Maxwell field equation for our spacetime yields
\begin{equation*}
\phi^{\prime\prime}+\left(\frac{2}{r}-\frac{\nu'}{2}-\frac{\lambda'}{2}\right)\phi'=
4\pi\zeta e^{\frac{\nu}{2}+\lambda},
\end{equation*}
where prime denotes differentiation with respect to $r$. Integration
of the above equation yields
\begin{equation*}
{\phi}'=\frac{e^{\frac{\nu+\lambda}{2}}q(r)}{r^{2}}.
\end{equation*}
Here $q(r)=4\pi\int^r_{0}{\zeta}e^{\frac{\lambda}{2}}{r^2}dr$
indicates total charge inside the sphere. The corresponding
Einstein-Maxwell field equations turn out to be
\begin{eqnarray}\label{4}
&&8{\pi}{\rho}+\frac{q^2}{r^{4}}
=\frac{e^{-\lambda}\lambda'}{r}-\frac{e^{-\lambda}-1}{r^{2}},\\\label{5}
&&8{\pi}{P_r }-\frac{q^2}{r^{4}}
=\frac{e^{-\lambda}\nu'}{r}+\frac{e^{-\lambda}-1}{r^{2}},\\\label{6}
&&8{\pi}{P_\bot }+\frac{q^2}{r^{4}}
=e^{-\lambda}\left(\frac{\nu''}{2}+\frac{\nu'^{2}}{4}-\frac{\nu'\lambda'}{4}
+\frac{\nu'}{2r}-\frac{\lambda'}{2r}\right).
\end{eqnarray}
The corresponding Misner-Sharp mass leads to \cite{24}
\begin{equation}\label{7}
m(r)=\frac{r}{2}\left(1-e^{-\lambda}\right)+\frac{q^2}{2r}.
\end{equation}
The conservation law, $T^{\mu}_{~\nu;\mu}+E^{\mu}_{~\nu;\mu} = 0$,
yields
\begin{equation}\label{8}
{P_r}'+\frac{\nu'}{2}\left(\rho+P_r\right)-\frac{2}{r}\left(P_\bot-P_r+\frac{qq'}
{8{\pi}r^{3}}\right)=0.
\end{equation}
This is termed as generalized TOV equation which represents charged
sphere in hydrostatic equilibrium. Using Eqs.(\ref{5}) and
(\ref{7}), we have
\begin{equation*}
\frac{\nu'}{2}=\frac{4{\pi}r^{4}{P_r}-q^{2}+{r}{m}}{r(r^{2}-2{r}{m}+q^{2})}.
\end{equation*}
Consequently, Eq.(\ref{8}) becomes
\begin{equation}\label{9}
{P_r}'+\frac{4{\pi}r^{4}{P_r}-q^{2}+{r}{m}}{r^{3}-2{r^2}{m}
+rq^{2}}(\rho+P_r)-\frac{2\Delta}{r}-\frac{qq'} {4{\pi}r^{4}}=0,
\end{equation}
where $\Delta=P_\bot-P_r$. The polytropic EoS has two possible cases
\cite{8}. The first case gives
\begin{equation}\label{10}
P_{r}=k\rho_{0}^{\gamma}=k\rho_{0}^{1+\frac{1}{n}},\quad\rho-\rho_{0}=nP_{r},
\end{equation}
where $\rho_{0}$ is baryonic density. The polytropic EoS for the
second case is
\begin{equation}\label{11}
P_{r}=k\rho^{\gamma}=k\rho^{1+\frac{1}{n}},\quad
\rho\left(1-k\rho_{0}^{\frac{1}{n}}\right)^{n}=\rho_{0},
\end{equation}
where $\rho_{0}$ is replaced by $\rho$.

In the following, we evaluate generalized TOV and mass equations for
the above two cases of polytropic EoS.

\subsection{Case 1}

We consider the first polytropic EoS (\ref{10}) and construct
generalized TOV equation. For this purpose, we take \cite{9}
\begin{eqnarray}\label{12}
P_{rc}=\alpha\rho_{c},\quad \xi=rA,\quad A^{2}=\frac{4
\pi{\rho_{c}}}{\alpha{(n+1)}}, \quad
\Phi_{0}^{n}=\frac{\rho_{0}}{\rho_{0c}},\quad
m(r)=\frac{4{\pi}\rho_{c}\eta(\xi)}{A^{3}},
\end{eqnarray}
where subscript $c$ indicates the value at the center,
$\alpha,~\xi,~\Phi_{0},~\eta$ are dimensionless variables and $A$ is
constant. Using the variables alongwith Eq.(\ref{10}) in (\ref{9}),
we obtain
\begin{eqnarray}\nonumber
&&\left\{\frac{1-\frac{2\eta\alpha(n+1)}{\xi}+\frac{4\pi
\rho_{c}{q}^{2}}{{\xi}^{2}\alpha(n+1)}}{\alpha(n+1)\Phi_{0}+1-
n\alpha}\right\}
\left(\xi^2\frac{d\Phi_{0}}{d\xi}-2\Phi_{0}^{-n}\frac{\alpha^4
(n+1)^2\xi^3\Delta+2\pi\alpha^2\rho_{c}^{2}q\frac{dq}{d\xi}}{\alpha^5\rho_{c}
(n+1)^3\xi^2}\right)\\\label{13} &&+\alpha\xi^{3}\Phi_{0}^{n+1}+\eta
-\frac{4\pi\rho_{c}q^2}{\xi\alpha^2 (n+1)^2}=0.
\end{eqnarray}
Differentiating Eq.(\ref{7}) and using (\ref{4}), we obtain the mass
equation as follows
\begin{equation}\label{14}
m'=4 \pi r^{2}\rho+\frac{qq'}{r}.
\end{equation}
Using Eqs.(\ref{10}) and (\ref{12}) in the above equation, we have
\begin{equation}\label{15}
\frac{d\eta}{d\xi}=\xi^{2}\Phi_{0}^{n}(1+n\alpha\Phi_{0}-n\alpha)
+\frac{4\pi\rho_{c}q}{\xi\alpha^2 (n+1)^2}\frac{dq}{d\xi}.
\end{equation}
The coupling of generalized TOV equation (\ref{13}) with mass
equation (\ref{15}) yields the Lane-Emden equation which provides
simple models for polytropes in hydrostatic equilibrium.

\subsection{Case 2}

In this case, we construct TOV equation for Eq.(\ref{11}) by taking
\begin{equation}\nonumber
\Phi^{n}=\rho/\rho_{c}.
\end{equation}
Consequently, Eq.(\ref{9}) turns out to be
\begin{eqnarray}\nonumber
&&\left\{\frac{1-\frac{2\eta\alpha(n+1)}{\xi}+\frac{4\pi
\rho_{c}{q}^{2}}{{\xi}^{2}\alpha(n+1)}}{1+\alpha\Phi}\right\}
\left(\xi^2\frac{d\Phi}{d\xi}-2\Phi^{-n}\frac{\alpha^4
(n+1)^2\xi^3\Delta+2\pi\alpha^2\rho_{c}^{2}q\frac{dq}{d\xi}}{\alpha^5\rho_{c}
(n+1)^3\xi^2}\right)\\\label{17} &&+\alpha\xi^{3}\Phi^{n+1}
+\eta-\frac{4\pi\rho_{c}q^2}{\xi\alpha^2 (n+1)^2}=0.
\end{eqnarray}
Using Eqs.(\ref{11}) and (\ref{12}) in (\ref{14}), it follows that
\begin{equation}\label{ccc}
\frac{d\eta}{d\xi}=\xi^{2}\Phi^{n}+\frac{4\pi\rho_{c}q}{\xi\alpha^2
(n+1)^2}\frac{dq}{d\xi}.
\end{equation}
Again, the coupling of above two equations provides the Lane-Emden
equation corresponding to this case. We see that Eqs.(\ref{13}),
(\ref{15}) and (\ref{17}), (\ref{ccc}) form two systems of
differential equations in three unknowns for case \textbf{1} and
\textbf{2}, respectively. In order to reduce one unknown, we
consider the following EoS \cite{c}
\begin{equation}\label{sss}
\Delta=\frac{B(4{\pi}r^{4}{P_r}-q^{2}+{r}{m})}{r^{2}-2{r}{m}
+q^{2}}(\rho+P_r),
\end{equation}
where $B$ is a constant.

\section{Cracking of Anisotropic Polytrope}

In astrophysical objects, the matter distribution may depart from
equilibrium state when perturbations are introduced. We analyze the
stability of polytropic compact object using the concept of
cracking. For this purpose, we use Eq.(\ref{sss}) in (\ref{9}) which
yields
\begin{equation}\label{ddd}
\mathcal{R}=\frac{dP_{r}}{dr}+\beta\left[\frac{4{\pi}r^{4}{P_r}
-q^{2}+{r}{m}}{r^{3}-2{r^2}{m}
+rq^{2}}\right](\rho+P_r)-\frac{qq'}{4{\pi}r^{4}},
\end{equation}
here $\beta=1-2B$ and $\mathcal{R}$ represents the force
distribution function. In order to observe cracking in our system,
we perturb matter variables for both cases of polytropic EoS through
a set of parameters ($k,\beta$) and ($n,\beta$).

\subsection{Perturbations in Case 1}

In this case, we construct force distribution function by perturbing
the energy density and anisotropy via $k~ \text{and}~ \beta$ as
follows
\begin{equation}\nonumber
k\rightarrow \tilde{k}=k+\delta k,\quad \beta\rightarrow
\tilde{\beta}=\beta+\delta \beta.
\end{equation}
After perturbation, Eq.(\ref{10}) takes the form
\begin{equation}
\tilde{P_{r}}=\tilde{k}\rho_{0}^{1+\frac{1}{n}}=\omega P_{r},\quad
\tilde{\rho}=\rho_{0}+n\omega P_{r},
\end{equation}
where $\omega=\frac{\tilde{k}}{k}$. Introducing the dimensionless
variable
$\hat{\tilde{\mathcal{R}}}=\frac{A}{4\pi\rho_{c}^{2}}\tilde{\mathcal{R}}$
and using the perturbed parameters alongwith Eq.(\ref{12}) in
(\ref{ddd}), we have
\begin{eqnarray}\nonumber
\hat{\tilde{\mathcal{R}}}&=&
\frac{\tilde{\beta}\Phi_{0}^{n}(1-n\alpha+\alpha\omega(n+1)\Phi_{0})}{\xi^{2}}
\left\{\frac{\tilde{\eta}+\alpha\omega\xi^{3}\Phi_{0}^{n+1}-
\frac{4\pi\rho_{c}q^2}{\xi\alpha^2
(n+1)^2}}{1-\frac{2\tilde{\eta}\alpha(n+1)}{\xi}+\frac{4\pi
\rho_{c}{q}^{2}}{{\xi}^{2}\alpha(n+1)}}\right\}\\\label{18}
&+&\omega\Phi_{0}^{n}\frac{d\Phi_{0}}{d\xi}-\frac{qq'}{4{\pi}r^{4}}.
\end{eqnarray}
In equilibrium state, the system has no perturbation which yields
$\hat{\tilde{\mathcal{R}}}(\xi,1+\delta\omega,
\beta+\delta\beta,\eta+\delta\eta)=0$. Applying Taylor's expansion,
we obtain
\begin{eqnarray}\nonumber
\mathcal{\delta\hat{R}}&=&\hat{\tilde{\mathcal{R}}}(\xi,1+\delta\omega,
\beta+\delta\beta,\eta+\delta\eta)=
\frac{\partial\hat{\tilde{\mathcal{R}}}}{\partial\omega}|_
{\omega=1, \tilde{\beta}=\beta,
\tilde{\eta}=\eta}\delta\omega\\\label{19}
&+&\frac{\partial\hat{\tilde{\mathcal{R}}}}{\partial\tilde{\beta}}|_
{\omega=1, \tilde{\beta}=\beta, \tilde{\eta}=\eta}\delta\beta +
\frac{\partial\hat{\tilde{\mathcal{R}}}}{\partial\tilde{\eta}}|_
{\omega=1, \tilde{\beta}=\beta, \tilde{\eta}=\eta}\delta \eta.
\end{eqnarray}
Using Eq.(\ref{18}), we evaluate
\begin{eqnarray}\nonumber
\frac{\partial\hat{\tilde{\mathcal{R}}}}{\partial\omega}|_
{\omega=1, \tilde{\beta}=\beta,
\tilde{\eta}=\eta}&=&\Phi_{0}^{n}\frac{d\Phi_{0}}{d\xi}+\frac{\beta\Phi_{0}^{n+1}\alpha}
{\xi^2\left(1-\frac{2\xi\eta\alpha(n+1)}{\xi}+\frac{4\pi
\rho_{c}{q}^{2}}{{\xi}^{2}\alpha(n+1)}\right)}\left[\alpha\xi^3\Phi_{0}^n
\left(2(n+1)\Phi_{0}\right.\right.\\\label{20}
&+&\left.\left.1-n\right)
+(n+1)\left\{\eta-\frac{4\pi\rho_{c}q^2}{\xi\alpha^2
(n+1)^2}\right\}\right],\\\nonumber
\frac{\partial\hat{\tilde{\mathcal{R}}}}{\partial\tilde{\beta}}|_
{\omega=1, \tilde{\beta}=\beta,
\tilde{\eta}=\eta}&=&\frac{\Phi_{0}^{n}}{\xi^2}\left\{\frac{1-n\alpha+\alpha\Phi_{0}(n+1)
} {1-\frac{2\eta\alpha(n+1)}{\xi}+\frac{4\pi
\rho_{c}{q}^{2}}{{\xi}^{2}\alpha(n+1)}}\right\}
\left[\eta+\alpha\xi^3\Phi_{0}^{n+1}\right.\\\label{21}
&-&\left.\frac{4\pi\rho_{c}q^2}{\xi\alpha^2
(n+1)^2}\right],\\\nonumber
\frac{\partial\hat{\tilde{\mathcal{R}}}}{\partial\tilde{\eta}}|_
{\omega=1, \tilde{\beta}=\beta,
\tilde{\eta}=\eta}&=&\frac{\beta\Phi_{0}^{n}}{\xi^2}\left\{\frac{1-n\alpha+
\alpha\Phi_{0}(n+1)}
{\left(1-\frac{2\eta\alpha(n+1)}{\xi}+\frac{4\pi
\rho_{c}{q}^{2}}{{\xi}^{2}\alpha(n+1)}\right)^2}\right\}\left[1+
2\alpha^2(n+1)\xi^2\Phi_{0}^{n+1}\right.\\\label{22}
&-&\left.\frac{4\pi \rho_{c}{q}^{2}}{{\xi}^{2}\alpha(n+1)}\right].
\end{eqnarray}

Making use of perturbed parameters in Eq.(\ref{15}), we obtain
\begin{equation*}
\tilde{\eta}=\int_{0}^{\xi}\left\{\hat{\xi}^2\Phi_{0}^{n}
(1-n\alpha+n\alpha\omega\Phi_{0})+\frac{4\pi\rho_{c}q}{\hat{\xi}\alpha^2
(n+1)^2}\frac{dq}{d\hat{\xi}}\right\}d\hat{\xi}.
\end{equation*}
Moreover, we have
\begin{equation}\label{23}
\delta\eta=\frac{\partial\tilde{\eta}}{\partial\omega}|_
{\omega=1}\delta\omega=n\alpha f_{1}(\xi)\delta\omega,
\end{equation}
where
\begin{equation}\nonumber
f_{1}(\xi)=\int_{0}^{\xi}\hat{\xi}^2\Phi_{0}^{n+1}d\hat{\xi}.
\end{equation}
Substituting Eqs.(\ref{20})-(\ref{23}) in (\ref{19}), it follows
that
\begin{eqnarray}\nonumber
\delta\mathcal{\hat{R}}_{1}&=&\Phi_{0}^{n}\left[\frac{d\Phi_{0}}{d\xi}
+\frac{\alpha\beta}{\xi^2\left(1-\frac{2\eta\alpha(n+1)}{\xi}+\frac{4\pi
\rho_{c}{q}^{2}}{{\xi}^{2}\alpha(n+1)}\right)}\left\{\alpha\xi^3\Phi_{0}^{n+1}
\left(2(n+1)\Phi_{0}\right.\right.\right.\\\nonumber
&+&\left.\left.\left.1-n\right)
+(n+1)\Phi_{0}\left\{\eta-\frac{4\pi\rho_{c}q^2}{\xi\alpha^2
(n+1)^2}\right\}+\left(1+
2\alpha^2(n+1)\xi^2\Phi_{0}^{n+1}\right.\right.\right.\\\nonumber
&-&\left.\left.\left.\frac{4\pi
\rho_{c}{q}^{2}}{{\xi}^{2}\alpha(n+1)}\right)
\left(\frac{1-n\alpha+\alpha\Phi_{0}(n+1)}
{1-\frac{2\eta\alpha(n+1)}{\xi}+\frac{4\pi
\rho_{c}{q}^{2}}{{\xi}^{2}\alpha(n+1)}}\right)nf_{1}(\xi)\right\}\right]\delta\omega\\\nonumber
&+&\frac{\Phi_{0}^{n}}{\xi^2}\left(1-n\alpha+\alpha(n+1)\Phi_{0}\right)
\left\{\frac{\eta+\alpha\xi^3\Phi_{0}^{n+1}-\frac{4\pi\rho_{c}q^2}{\xi\alpha^2
(n+1)^2}}{1-\frac{2\eta\alpha(n+1)}{\xi}+\frac{4\pi
\rho_{c}{q}^{2}}{{\xi}^{2}\alpha(n+1)}}\right\}\delta\beta.
\end{eqnarray}
Using the variables $x=\frac{\xi}{\bar{A}},~
\bar{A}=r_{\Sigma}A=\xi_{\Sigma}$, the above equation yields
\begin{eqnarray}\nonumber
\delta\mathcal{\hat{R}}_{1}&=&\Phi_{0}^{n}\left[\frac{d\Phi_{0}}{\bar{A}dx}
+\frac{\alpha\beta}{\bar{A}^2x^2G_{1}}\left\{\alpha\bar{A}^3x^3\Phi_{0}^{n+1}
\left(2(n+1)\Phi_{0}+1-n\right)+(n+1)\Phi_{0}\left\{\eta\right.\right.\right.\\\nonumber
&-&\left.\left.\left.\frac{4\pi\rho_{c}q^2}{\bar{A}x\alpha^2
(n+1)^2}\right\}+\frac{G_{2}}{G_{1}}\left(1+
2\alpha^2(n+1)x^2\bar{A}^2\Phi_{0}^{n+1}-\frac{4\pi
\rho_{c}{q}^{2}}{x^{2}\bar{A}^2\alpha(n+1)}\right)\right.\right.\\\label{a}
&\times&\left.\left. nf_{1}(x)\right\}\right]\delta\omega
+\frac{\Phi_{0}^{n}G_{2}G_{3}}{x^2\bar{A}^{2}G_{1}}\delta\beta.
\end{eqnarray}
where
\begin{eqnarray}\nonumber
G_{1}&=&1-\frac{2\alpha(n+1)\eta}{\bar{A}x}+\frac{4\pi
\rho_{c}{q}^{2}}{\bar{A}^{2}x^2\alpha(n+1)},\\\nonumber
G_{2}&=&1-n\alpha+\alpha\Phi_{0}(n+1),\\\nonumber
G_{3}&=&\eta+\alpha
x^3\bar{A}^3\Phi_{0}^{n+1}-\frac{4\pi\rho_{c}q^2}{x\bar{A}\alpha^2
(n+1)^2}.
\end{eqnarray}
In cracking, $\delta\hat{\mathcal{R}}>0$ inside the sphere while
$\delta\hat{\mathcal{R}}<0$ for outer regions, so
$\delta\hat{\mathcal{R}}=0$ for some value of $\xi$. This condition
implies that
\begin{equation*}
\delta\omega=-\frac{\delta\beta}{\Gamma},
\end{equation*}
where
$\Gamma=\frac{\frac{\partial\hat{\tilde{\mathcal{R}}}}{\partial\omega}+
\frac{\partial\hat{\tilde{\mathcal{R}}}}{\partial\tilde{\eta}}f_{1}(\xi)}
{\frac{\partial\hat{\tilde{\mathcal{R}}}}{\partial\tilde{\beta}}}|_
{\omega=1, \tilde{\beta}=\beta, \tilde{\eta}=\eta}$.

We study a phenomenon of cracking by perturbing energy density and
anisotropy via parameters $n~\text{and}~\beta$ as follows
\begin{equation}\nonumber
n\rightarrow\tilde{n}=n+\delta n,\quad
\beta\rightarrow\tilde{\beta}=\beta+\delta\beta.
\end{equation}
Equation (\ref{ddd}) in terms of perturbed parameters can be written
as
\begin{eqnarray}\nonumber
\hat{\tilde{\mathcal{R}}}&=&\Phi_{0}^{\tilde{n}}\frac{d\Phi_{0}}{d\xi}+
\frac{\tilde{\beta}\Phi_{0}^{\tilde{n}}\left(1-\tilde{n}\alpha+\alpha
(\tilde{n}+1)\Phi_{0}\right)}{\xi^{2}}
\left\{\frac{\tilde{\eta}+\alpha\xi^{3}\Phi_{0}^{\tilde{n}+1}
-\frac{4\pi\rho_{c}q^2}{\xi\alpha^2
(\tilde{n}+1)^2}}{1-\frac{2\tilde{\eta}\alpha(\tilde{n}+1)}{\xi}+\frac{4\pi
\rho_{c}{q}^{2}}{{\xi}^{2}\alpha(\tilde{n}+1)}}\right\}\\\label{25}
&-&\frac{qq'}{4{\pi}r^{4}}.
\end{eqnarray}
In this scheme, Taylor's expansion yields
\begin{eqnarray}\nonumber
\delta\hat{\mathcal{R}}=\frac{\partial\hat{\tilde{\mathcal{R}}}}{\partial\tilde{n}}|_
{\tilde{n}=n, \tilde{\beta}=\beta, \tilde{\eta}=\eta}\delta n
+\frac{\partial\hat{\tilde{\mathcal{R}}}}{\partial\tilde{\eta}}|_
{\tilde{n}=n, \tilde{\beta}=\beta, \tilde{\eta}=\eta}\delta
\eta+\frac{\partial\hat{\tilde{\mathcal{R}}}}{\partial\tilde{\beta}}|_
{\tilde{n}=n, \tilde{\beta}=\beta, \tilde{\eta}=\eta}\delta\beta.
\end{eqnarray}
Making use of Eq.(\ref{25}), the above equation turns out to be
\begin{eqnarray}\nonumber
\delta\mathcal{\hat{R}}_{2}&=&\Phi_{0}^{n}\left[\frac{\ln\Phi_{0}}{\bar{A}}\frac{d\Phi_{0}}
{dx}+\frac{\beta}{\bar{A}^{2}x^2G_{1}}\left\{\ln\Phi_{0}G_{2}G_{3}
+\alpha(\Phi_{0}-1)G_{3}+G_{2}\left(\alpha
A^{3}x^3\right.\right.\right.\\\nonumber
&\times&\left.\left.\left.\Phi_{0}^{n+1}\ln\Phi_{0}
+\frac{8\pi\rho_{c}q^2}{\bar{A}x\alpha^{2}(n+1)^3}\right)+\frac{G_{2}G_{3}}{G_{1}}
\left(\frac{2\eta\alpha}{\bar{A}x}+
\frac{4\pi\rho_{c}q^2}{\bar{A}^2x^2\alpha(n+1)^2}\right)\right.\right.\\\label{b}
&+&\left.\left.\frac{G_{2}}{G_{1}}\left(G_{1} +
\frac{2\alpha(n+1)}{\bar{A}x}G_{3}\right)
f_{2}(x)\right\}\right]\delta
n+\frac{\Phi_{0}^{n}}{\bar{A}^2x^2}\frac{G_{2}G_{3}}{G_{1}}\delta\beta,
\end{eqnarray}
where
\begin{equation}\label{bb}
f_{2}(x)=\int_{0}^{x}\bar{A}^3\hat{x}^2\Phi_{0}^{n}\left\{
\alpha(\Phi_{0}^{1-n}-1)+(1-n\alpha)\ln\Phi_{0}\right\}d\hat{x}.
\end{equation}
Again, for cracking to occur, $\delta\hat{\mathcal{R}}=0$ implying
that
\begin{equation}\nonumber
\delta n=-\frac{\delta\beta}{\Gamma},
\end{equation}
where
$\Gamma=\frac{\frac{\partial\hat{\tilde{\mathcal{R}}}}{\partial\tilde{n}}+
\frac{\partial\hat{\tilde{\mathcal{R}}}}{\partial\tilde{\eta}}f_{2}(\xi)}
{\frac{\partial\hat{\tilde{\mathcal{R}}}}{\partial\tilde{\beta}}}|_
{\tilde{n}=n, \tilde{\beta}=\beta, \tilde{\eta}=\eta}$.

\subsection{Perturbations in Case 2}

Here, we develop force distribution function for the second kind of
polytropic EoS using perturbed parameters ($k,\beta$). After
perturbation, the energy density takes the form
\begin{equation}\nonumber
\tilde{\rho}=\rho+\delta\rho=\rho+\frac{\partial\tilde{\rho}}
{\partial\omega}|_{\omega=1},
\end{equation}
which yields
\begin{equation}\nonumber
\tilde{\rho}=\rho+nP_{r}(1-\omega),
\end{equation}
where we have used $\omega=1+\delta\omega$. Using perturbed
parameters in Eq.(\ref{ddd}), we have
\begin{eqnarray}\nonumber
\hat{\tilde{\mathcal{R}}}&=&\omega\Phi^{n}\frac{d\Phi}{d\xi}+
\frac{\tilde{\beta}\Phi^{n}\left[1+\alpha\Phi\{n+\omega(1-n)\}\right]}
{\xi^{2}\left(1-\frac{2\eta\alpha(n+1)}{\xi}+\frac{4\pi
\rho_{c}{q}^{2}}{{\xi}^{2}\alpha(n+1)}\right)}
\left\{\tilde{\eta}+\alpha\omega\xi^{3}\Phi^{n+1}\right.\\\nonumber
&-&\left.\frac{4\pi\rho_{c}q^2}{\xi\alpha^2
(n+1)^2}\right\}-\frac{qq'}{4{\pi}r^{4}}.
\end{eqnarray}
Applying Taylor's expansion on the above equation, it follows that
\begin{eqnarray}\nonumber
\delta\mathcal{\hat{R}}_{3}&=&\Phi^{n}\left[\frac{d\Phi}{\bar{A}dx}
+\frac{\alpha\beta}{\bar{A}^2x^2G_{1}}\left\{\Phi(1-n)\left(\eta-\frac{4\pi\rho_{c}q^2}{x\bar{A}\alpha^2
(n+1)^2}\right)+ x^3\bar{A}^3\alpha(1-n)\right.\right.\\\nonumber
&\times&\left.\left.\Phi^{n+2} -n(1+\alpha\Phi)
\left(\frac{1+2\alpha^2\bar{A}^2x^2(n+1)\Phi^{n+1}} {G_{1}}\right)
f_{3}(x)\right\}\right]\delta\omega\\\label{c}
&+&\frac{\Phi^{n}}{x^2\bar{A}^2}\left(1+\alpha\Phi\right)
\frac{G_{3}}{G_{1}}\delta\beta,
\end{eqnarray}
where $f_{3}(x)=\int_{0}^{x}\bar{A}^3\hat{x}^2\Phi^{n+1}d\hat{x}$.
Similarly, perturbation of $n~ \text{and}~ \beta$ leads to
\begin{eqnarray}\nonumber
\delta\mathcal{\hat{R}}_{4}&=&\Phi^{n}\left[\frac{\ln\Phi}{\bar{A}}\frac{d\Phi}
{dx}+\frac{\beta}{\bar{A}^{2}x^2}\left\{\frac{\ln\Phi
G_{4}G_{5}}{G_{1}} +\frac{G_{5}}{G_{1}}\left(\alpha
A^{3}x^3\Phi^{n+1}\ln\Phi\right.\right.\right.\\\nonumber
&+&\left.\left.\left.
\frac{8\pi\rho_{c}q^2}{\bar{A}x\alpha^{2}(n+1)^3}\right)
-\frac{G_{4}G_{5}}{G_{1}^2} \left(-\frac{2\eta\alpha}{\bar{A}x}-
\frac{4\pi\rho_{c}q^2}{\bar{A}^2x^2\alpha(n+1)^2}\right)\right.\right.\\\label{d}
&+&\left.\left.\frac{G_{5}}{G_{1}^{2}}\left(G_{1}+
\frac{2\alpha(n+1)}{\bar{A}x}\bar{G}_{4}\right)f_{2}(x)\right\}\right]\delta
n+\frac{\Phi^{n}}{\bar{A}^2x^2}\frac{G_{4}G_{5}}{G_{1}}\delta\beta,
\end{eqnarray}
where
\begin{eqnarray}\nonumber
G_{4}&=&\eta+\alpha
x^3\bar{A}^3\Phi^{n+1}-\frac{4\pi\rho_{c}q^2}{x\bar{A}\alpha^2
(n+1)^2},\quad G_{5}=1+\alpha\Phi,\\\nonumber
f_{4}(x)&=&\int_{0}^{x}\bar{A}^{3}\hat{x}^2\Phi^{n}\ln\Phi d\hat{x}.
\end{eqnarray}

Now we analyze the occurrence of cracking in the polytropic models
through numerical method. We examine the charged compact models for
both cases of polytropic EoS with perturbations through ($k,\beta$)
and ($n,\beta$). Firstly, we evaluate $\Phi_{0}~,\Phi~ \text{and}~
\eta$ for both cases by integrating Eqs.(\ref{13}), (\ref{15}),
(\ref{17}) and (\ref{ccc}) with boundary conditions \cite{9}
\begin{equation}\nonumber
\eta(0)=0,\quad
\Phi_{0}(0)=1,\quad\Phi_{0}(\xi_{\Sigma})=0,\quad\Phi(0)=1,\quad
\Phi(\xi_{\Sigma})=0,
\end{equation}
and use these results to plot force distribution functions. The
graphical behavior of force distribution function for the case
\textbf{1} is shown in Figure \textbf{1}. The left graph shows the
behavior of $\delta\mathcal{\hat{R}}_{1}$ indicating that for all
considered values of $q$, it is positive in the inner regions and
becomes negative for the outer ones thus ensuring the occurrence of
strongest cracking in the corresponding model. The right graph is
plotted for $\delta\mathcal{\hat{R}}_{2}$ which shows that there is
neither cracking nor overturning for all values of $q$. Thus, the
presence of charge in polytropes leads to stable models.

For the case \textbf{2}, the plots of force function in
Eqs.(\ref{c}) and (\ref{d}) for different values of the parameters
are shown in Figure \textbf{2}. The left graph is plotted for
$\delta\mathcal{\hat{R}}_{3}$ which indicates stable behavior for
$q=0.2M_{0}$ while the strongest overturning appears for other two
values of $q$. The graphical analysis of
$\delta\mathcal{\hat{R}}_{4}$ for different values of charge shows
stable configuration as shown in Figures \textbf{2} (right) and
\textbf{3}. It is observed that within uncharged matter distribution
both cracking and overturning appear. However, the inclusion of
charge in matter configurations leads to stability of spherically
symmetric polytrope.
\begin{figure}\centering
\epsfig{file=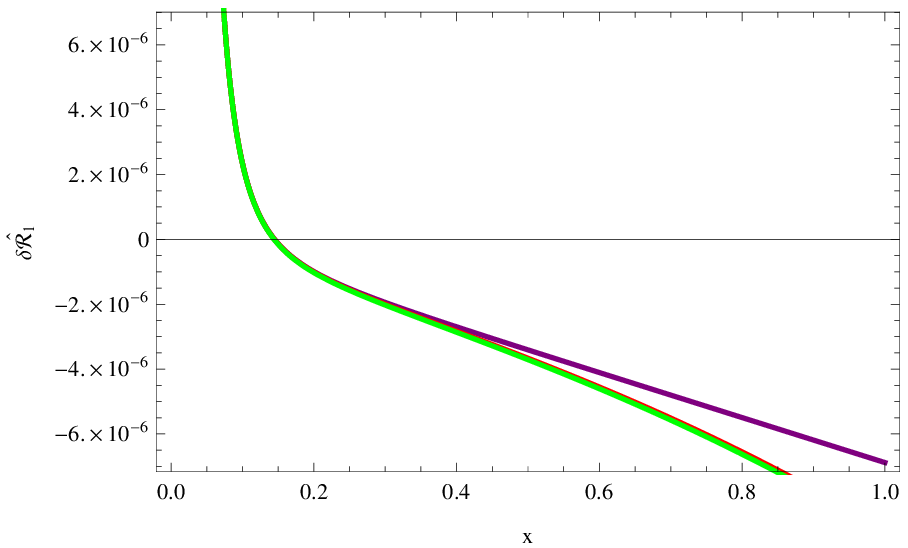,width=0.45\linewidth}
\epsfig{file=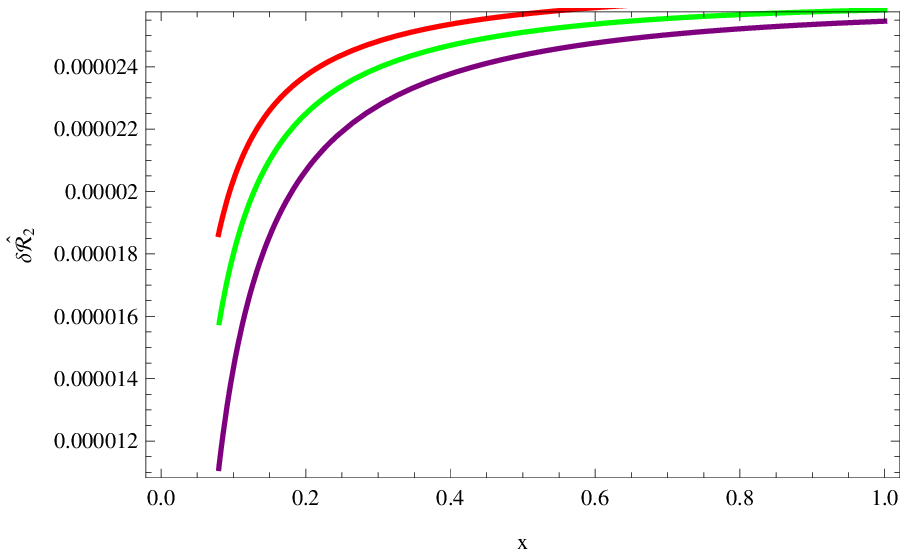,width=0.45\linewidth} \caption{Left plot for
$\delta\hat{\mathcal{R}_{1}}$ versus $x$ with
$n=1,\alpha=0.85,\beta=1.5,\Gamma=1.6,q=0.2M_{0}$ (purple),
$q=0.4M_{0}$ (red), $q=0.64M_{0}$ (green). Right plot for
$\delta\hat{\mathcal{R}_{2}}$ versus $x$ with
$n=1.5,\alpha=0.90,\beta=0.5,\Gamma=1.4,q=0.2M_{0}$ (purple),
$q=0.4M_{0}$ (red), $q=0.64M_{0}$ (green).}
\end{figure}
\begin{figure}\centering
\epsfig{file=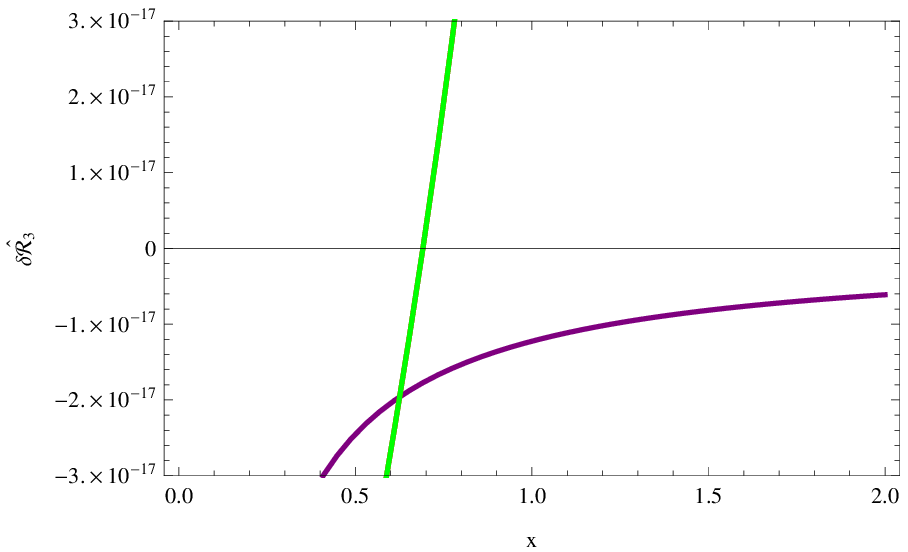,width=0.45\linewidth}
\epsfig{file=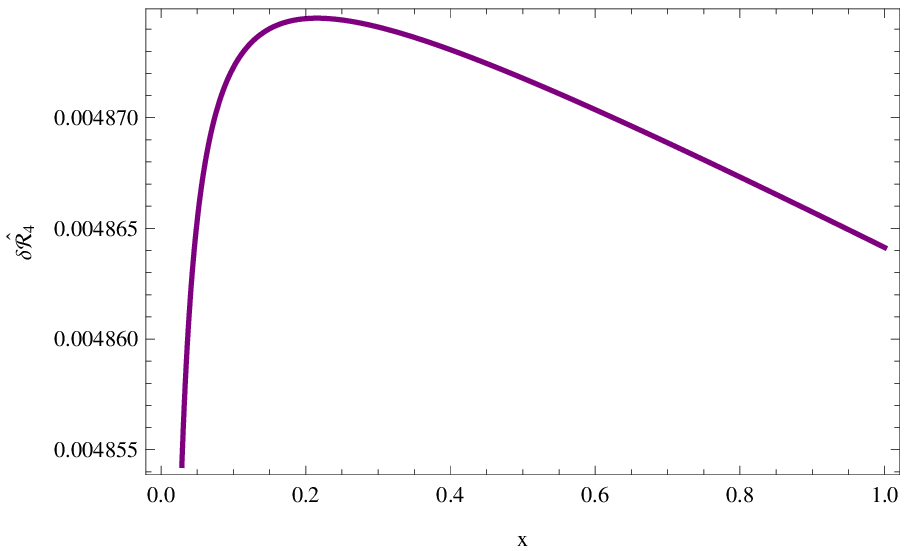,width=0.45\linewidth} \caption{Left plot for
$\delta\hat{\mathcal{R}_{3}}$ versus $x$ with
$n=0.5,\alpha=0.98,\beta=1.5,\Gamma=0.6,q=0.2M_{0}$ (purple),
$q=0.4M_{0}$ (red), $q=0.64M_{0}$ (green). Right plot for
$\delta\hat{\mathcal{R}_{4}}$ versus $x$ with
$n=1.5,\alpha=0.7,\beta=1.5,\Gamma=-0.5,q=0.2M_{0}$.}
\end{figure}
\begin{figure}\centering
\epsfig{file=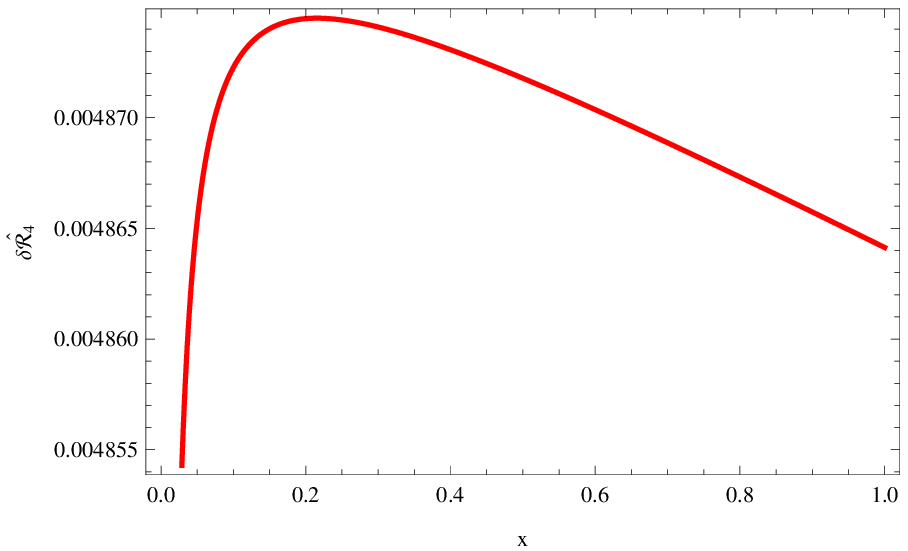,width=0.45\linewidth}
\epsfig{file=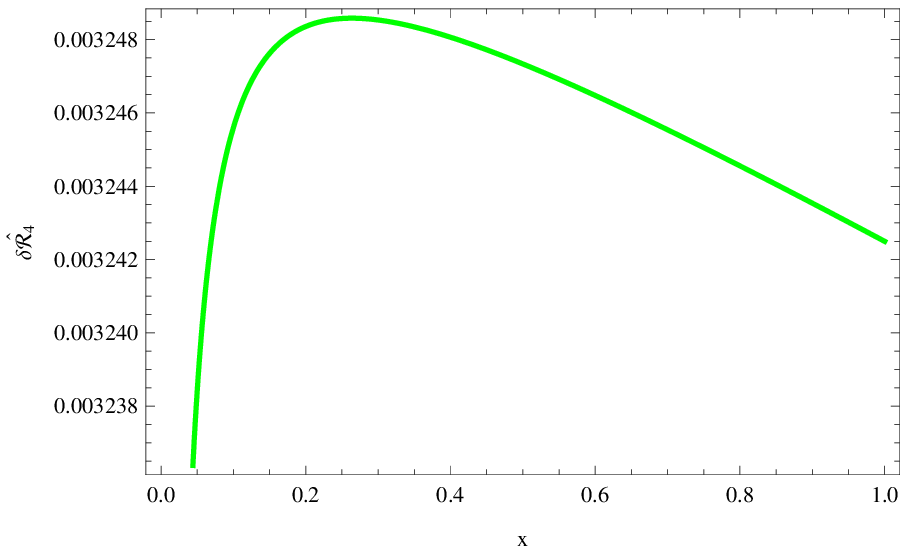,width=0.45\linewidth} \caption{Plot for
$\delta\hat{\mathcal{R}_{4}}$ versus $x$ with
$n=1.5,\alpha=0.7,\beta=1.5,\Gamma=-0.5$. Left plot for
$q=0.4M_{0}$. Right plot for $q=0.64M_{0}$.}
\end{figure}

\section{Concluding Remarks}

The stability analysis of stellar objects is an interesting area of
research. The concept of cracking refers to the appearance of total
radial forces of different signs within the matter distribution. If
there is no cracking (or overturning) within a particular
configuration, then the system is not completely stable since other
types of perturbation can lead to its cracking, overturning,
expansion or collapse. We have considered spherically symmetric star
with charged anisotropic matter distribution satisfying polytropic
EoS and constructed the corresponding Einstein-Maxwell equations. We
have considered two cases of polytropic EoS and formulated TOV as
well as mass equations in terms of dimensionless variables for each
case. The coupling of these two equations represent charged
polytrope in hydrostatic equilibrium.

In order to observe cracking, perturbation is a necessary ingredient
to take out system from equilibrium state. We have perturbed energy
density and pressure anisotropy of a system in two ways. Firstly, we
have introduced perturbations through parameters ($k,\beta$) and
constructed the force distribution function
$(\delta\mathcal{\hat{R}}_{1})$ in terms of perturbed parameters
describing total radial forces present in a system. The graphical
analysis of $\delta\mathcal{\hat{R}}_{1}$ indicates the appearance
of cracking for all choices of parameters thus leading to unstable
configurations for this case. Secondly, we have used ($n,\beta$) as
perturbation parameters and constructed
$\delta\mathcal{\hat{R}}_{2}$. It is found that the resulting models
are stable towards perturbations.

We have followed the same procedure for the case \textbf{2} of
polytropic EoS and constructed $\delta\mathcal{\hat{R}}_{3}~
\text{as well as} ~\delta\mathcal{\hat{R}}_{4}$. It is found that
polytropic models are unstable for perturbation in ($k,\beta$),
while the perturbation of ($n,\beta$) leads to stable matter
configuration representing relativistic polytrope. The stability of
compact objects depends upon the choice of EoS. It was found that
spherically symmetric charged compact objects with quadratic EoS
\cite{4s} lead to stable models while the linear EoS \cite{sss}
provides unstable configurations. For uncharged spherical
anisotropic polytropes, both cracking and overturning appear for a
wide range of parameters under ($k,\beta$) as well as ($n,\beta$)
perturbations \cite{c}. We have observed that charged matter
distribution leads to stable configurations for ($n,\beta$)
perturbation while polytropic models remain unstable by perturbing
($k,\beta$). We conclude that that the inclusion of charge in
anisotropic fluid distribution has a dominant effect on polytropes
which leads to stable models.

\end{document}